\documentclass[12pt]{article}
\usepackage{times}
\input{psfig.sty}
\usepackage{amsmath}
\begin{document}
\pagestyle{plain}
\hsize = 6.5 in
\vsize = 8.5 in
\hoffset = -0.5 in
\voffset = -0.5 in
\baselineskip = 0.29 in
\def\vA{\mbox{\boldmath$A$}}
\def\vF{\mbox{\boldmath$F$}}
\def\vecJ{\mbox{\boldmath$J$}}
\def\vQ{\mbox{\boldmath$Q$}}
\def\vX{\mbox{\boldmath$X$}}
\def\vecPi{\mbox{\boldmath$\Pi$}}
\def\rvx{{\bf x}}
\def\rvz{{\bf z}}
\def\vF{{\bf F}}
\def\vG{{\bf G}}
\def\vJ{{\bf J}}
\def\vS{{\bf S}}
\def\vK{{\bf K}}
\def\rvxi{\hbox{\boldmath $\xi$}}
\def\vPi{\hbox{\boldmath $\Pi$}}

\begin{center}

\vskip 0.5cm
{\Large\bf Nonequilibrium Potential Function of Chemically\\[10pt]
Driven Single Macromolecules via\\[10pt]
Jarzynski-Type Log-Mean-Exponential Work}

\vskip 0.5 true cm 
Hong Qian
\vskip 0.5 true cm

Departments of Applied Mathematics\\
University of Washington, Seattle, WA 98195
\end{center}

\vskip 0.3cm
\centerline{\today}
\vskip 1. true cm

\begin{quote}
Applying the method from recently developed fluctuation
theorems to the stochastic dynamics of single macromolecules
in ambient fluid at constant temperature, we establish two 
Jarzynski-type equalities: (1) between the log-mean-exponential 
(LME) of the irreversible heat dissiption of a driven molecule in 
nonequilibrium steady-state (NESS) and $\ln P^{ness}(x)$, 
and (2) between the LME of the work done by the 
internal force of the molecule and nonequilibrium 
chemical potential function $\mu^{ness}(x)$  
$\equiv U(x)+k_BT\ln P^{ness}(x)$, where $P^{ness}(x)$ is 
the NESS probability density in the phase space of the
macromolecule and $U(x)$ is its internal potential function.  
$\Psi$ = $\int\mu^{ness}(x)P^{ness}(x)dx$ 
is shown to be a nonequilibrium generalization of the 
Helmholtz free energy and 
$\Delta\Psi$ = $\Delta U-T\Delta S$ for 
nonequilibrium processes, where $S$ 
$=-k_B\int P(x)\ln P(x)dx$ is 
the Gibbs entropy associated with $P(x)$.  LME of heat 
dissipation generalizes the concept of entropy, and the 
equalities define thermodynamic potential functions for 
open systems far from equilibrium.

\vskip 0.3cm

\end{quote}

\newpage

\vskip 0.5in\noindent
{\bf Introduction}
\vskip 0.3cm

	One of the important concepts in equilibrium thermodynamics 
is chemical potential.  In an equilibrium, the chemical potentials 
of all participating chemical components are equal.  Chemical 
potential also plays an important role in isothermal, driven, 
single molecular motor systems in nonequilibrium steady-state 
(NESS) \cite{qianpre04}.  Potential functions for stochastic 
systems has been investigated previous by Graham and Haken 
\cite{graham1,graham3,graham4} and recently by 
Ao and coworkers \cite{aoping1,aoping2}.  In all the above works, 
the logarithm of the probability distribution in a NESS that is
analogous to Boltzmann's entropy has been identified as a 
nonequilibrium potential function (NePf). 

	In classical mechanics and field theory, introducing a 
potential function is always associated with integration of a 
conservative force field.  Thus, for the NePfs introduced above, 
one necessarily seeks an ``integral interpretation''.  This is 
the objective of the present work which shows that, in 
a statistical sense, the integration is provided by the 
recent developed fluctuation theorems (FTs) and 
Jarzynski-type equalities in NESS \cite{jarzy,sasa}.  
FTs were first developed in the context of microcanonical chaotic 
dynamical systems \cite{gallavotti1,gallavotti2}.  It has also 
been studied in parallel for stochastic canonical dynamics 
\cite{ft1,ft2}.  In this paper, we first present our results via 
a simple example: a 3-state, driven cyclic reaction widely found 
in biochemistry.  Then the more general derivations based 
on the polymer representation of a macromolecule are given.  
Useful mathematics can be found in the Appendices.

\vskip 0.5cm\noindent
{\bf NePf and log-mean-exponential of heat and work: an example}  
\vskip 0.3cm 

	Fig. 1 shows a driven cyclic reaction of a single 
enzyme in aqueous solution at constant temperature $T$, 
sustained by constant concentrations of cofactors $D$ and $E$.   
This can be rigorously accomplished in a test tube with a feedback 
regenerating system, or approximately with a large excess of 
the cofactors \cite{xielu}.  In a living cell, such NESS reactions, 
e.g., kinase catalyzed phosphorylations and GAP-activated GTP
hydrolysis \cite{qianbpc03}, are ultimately maintained by the 
working of mitochondria which constantly regenerate ATP from ADP.

\begin{figure}[b]
\[
\psfig{figure=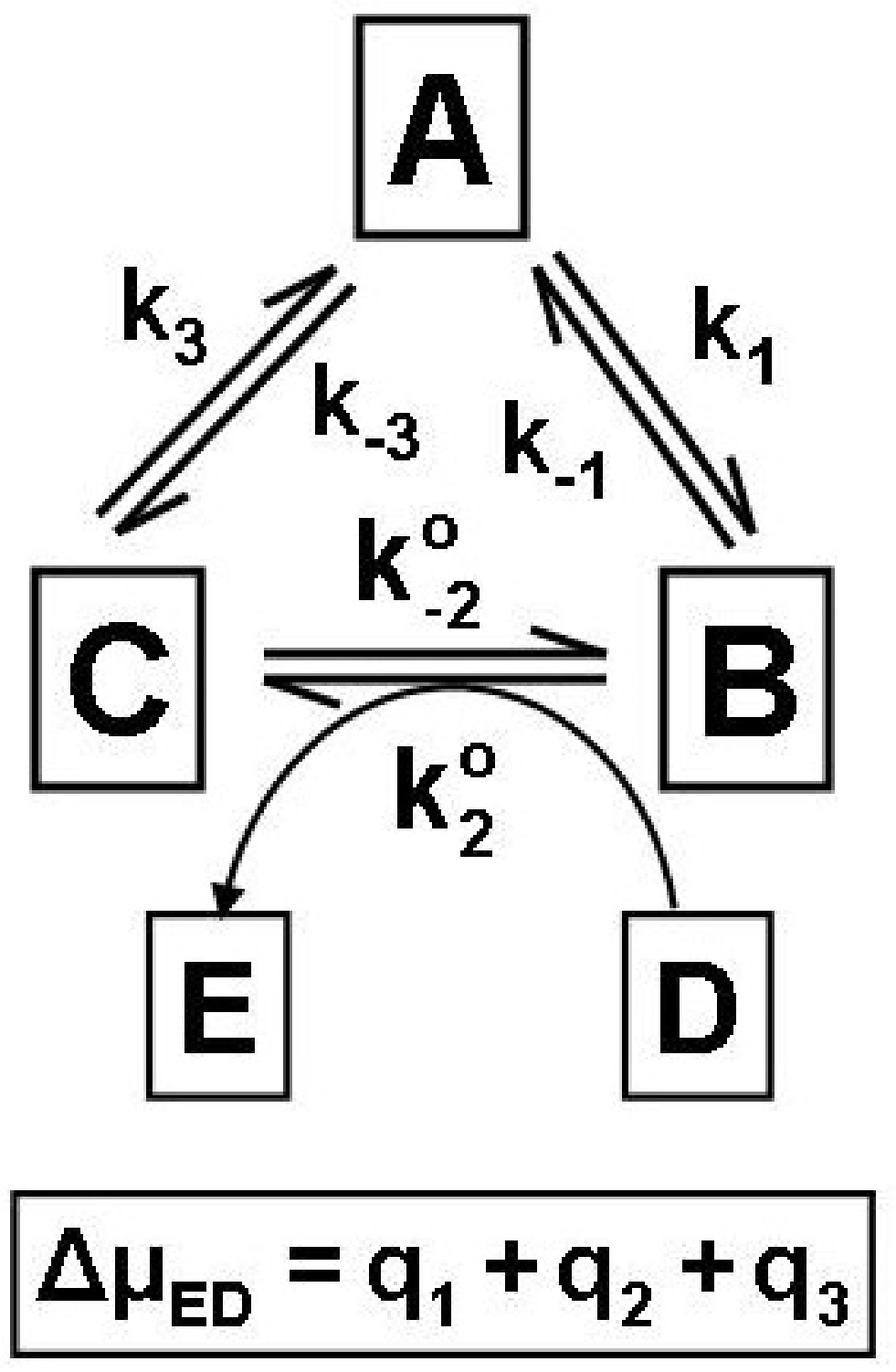,%
width=2.in,height=3.2in,%
bbllx=0.0in,bblly=0.in,%
bburx=4.in,bbury=6.25in}
\]
\caption{A single macromolecule can be in states $A$, $B$, and 
$C$.  The transition from $B$ to $C$ is accompanied by the 
turnover of a cofactor between $D$ and $E$ via the second-order 
$k_2^o$ and $k_{-2}^o$.  We denote pseudo-first-order rate 
constants $k_2^oc_D=k_2$ and $k_{-2}^oc_E=k_{-2}$.  The total
chemical driving force for the cyclic reaction is $\Delta\mu_{ED}$
given in Eq. \ref{necp43sm}. $q_1$ = $k_BT\ln(p_Ak_1/p_Bk_{-1})$,
$q_2$ = $k_BT\ln(p_Bk_2/p_Ck_{-2})$, and $q_1$ = 
$k_BT\ln(p_Ck_3/p_Ak_{-3})$, are the heat dissipation associated 
with steps 1, 2, and 3, respectively.  The equality in the box
reflects the conservation of energy.}
\end{figure}

	The chemical potential for a species $X$ ($=D$ or $E$) 
is $\mu_X$ $=\mu_X^o+k_BT\ln c_X$ where $c_X$ is the 
concentration for the species $X$. Then it is easy to show that
\begin{equation}
	\Delta\mu_{ED} = \mu_D-\mu_E = k_BT\ln
		\frac{k_1k_2^ok_3c_D}{k_{-1}k_{-2}^ok_{-3}c_E}.
\label{necp43sm}
\end{equation}
The chemical potentials for the single 
macromolecule are $\mu_X$ $=\mu_X^o+k_BT\ln p_X$ where $p_X$
is the probability of the macromolecule being in state $X$ 
($=A,B,$ or $C$).  If the reaction is in a closed vessel, it 
eventually reaches an equilibrium and $\Delta\mu_{ED}=0$ $\Rightarrow$ 
$c_E^{eq}/c_D^{eq}$ $=\frac{k_{-1}k_{-2}^ok_{-3}}{k_1k_2^ok_3}$
$=e^{-\Delta\mu_{ED}^0/k_BT}$.  

	However, if one controls the reaction such that both 
concentrations of $D$ and $E$ are maintained at some constant 
values, say $c_D$ and $c_E$, then Eq. \ref{necp43sm} gives the 
amount of work done to the system in order to sustain the NESS.  
The NESS flux 
\begin{eqnarray}
      J &=& k_1p_A^{ness}-k_{-1}p_B^{ness} 
	= k_2p_B^{ness}-k_{-2}p_C^{ness}
	= k_3p_C^{ness}-k_{-3}p_A^{ness} 
\nonumber\\[10pt]
	&=& \frac{k_1k_2^ok_3c_D-k_{-1}k_{-2}^ok_{-3}c_E}
		{k_1k_2+k_1k_3+k_2k_3+k_1k_{-2}+k_2k_{-3}
		+k_3k_1+k_{-1}k_{-2}+k_{-1}k_{-3}+k_{-2}k_{-3}}.
\\
	&&
\nonumber
\end{eqnarray}
We thus have $J\times\Delta\mu_{ED}$, the work done by the external 
agent per unit time, $\ge 0$ and being zero if and only if
when $\Delta\mu_{ED}=J=0$. The analogy between this and power =
current $\times$ voltage, from electrical circuits, should be 
noted.  The work also is equal to the sum of the 
dissipated heat from each every isothermal reaction: 
$\Delta\mu_{ED}$ $=q_1+q_2+q_3$ (see Fig. 1).  These  results reflect 
both the First and Second Laws of Thermodynamics in a single 
macromolecule far-from-equilibrium \cite{qianjpcm05}.  By far from 
equilibrium, we mean in general for a reversible reaction
$X \overset{k_+}{\underset{k_-}{\rightleftharpoons}}Y$ 
in a NESS network, $J$ = $k_+c_X-k_-c_Y=$ 
$J_+-J_-$ and $\Delta\mu_{YX}$ = $k_BT\ln (k_+c_X/k_-c_Y)=$ 
$k_BT\ln(J_+/J_-)$ are not linearly related \cite{hdf2}.  
However, if $J\ll J_+,J_-$, then one has $\Delta\mu\propto J$.  

	So far, all the materials are from standard 
chemical thermodynamics. (Though seldomly it has been presented 
in this way.)  To introduce the new results from the recent 
work on FTs \cite{ft1,ft2,crooks}, we first define the 
irreversible heat dissipation following a stochastic trajectory 
in a NESS, $Q_{irr}(t)$.  Let the single macromolecule having 
the trajectory $x_0, x_1, ..., x_n$ ($x_i = A,B,$ or $C$) with 
transition from state $x_{i-1}$ to state $x_i$ at $t_i$ 
($t_0=0 <t_1 < ... < t_n < t$), and defining
\begin{equation}
         Q_{irr}(t) = -\sum_{i=1}^n \Delta\mu_{x_{i-1}x_i}
\end{equation}
where $\Delta\mu_{AB}=q_1$, $\Delta\mu_{BC}=q_2$,  and
$\Delta\mu_{CA}=q_3$, all positive.  $Q_{irr}(t)$ is called
irreversible heat since the dissipation here are truly 
irreversible: $Q_{irr}(t)\equiv 0$ in an equilibrium steady-state. 
There is, however, a fluctuating reversible heat exchange between the 
system and its environment even in equilibrium.  The latter
term is not contained in the $Q_{irr}(t)$.  In a NESS, 
for each every reaction
$\Delta\mu$ $=\Delta h^o-T\Delta s^o-T\Delta s_{concentration}$, 
hence the total heat dissipation $=\Delta\mu-\Delta h^o=-T\Delta s$,
where entropy change $\Delta s$ includes macromolecular structural 
entropy change $\Delta s^o$ and concentration entropy change 
$\Delta s_{concentration}$.  According to the recent work of U. 
Seifert \cite{seifert}, $\Delta\mu/T$ is the total (irreversible) 
entropy production, $\Delta s$ is the entropy change, and 
$\Delta h^o/T$ is the exchanged heat which can be positive or
negative.  If the reaction is in an isothermal equilibrium, 
$\Delta\mu=0$ and fluctuation in heat exchange is equal to the
enthalpy fluctuation along a stochastic trajectory.
This exact compensation between fluctuating entropy and enthalpy
was first discussed in \cite{hdf3}.  It is a direct consequence of
microscopic energy conservation.   
	
	Applying the methods developed for FTs 
\cite{crooks,seifert,jqz,gaspard}, it can be shown that 
\begin{equation}
	\ln p_X^{ness} 
	= \ln\Big\langle\delta_{X,x_n}
				e^{-Q_{irr}(t)/k_BT}\Big\rangle
\label{logP}
\end{equation}
where $\langle\cdots\rangle$ is ensemble average over all 
the possible trajectories with initial distribution $p^{ness}$ 
for $x_0$, and $\delta_{i,j}$ is the Kronecker-delta function 
for discrete $i$ and $j$.  This is a new result.  In the light 
of Jarzynski's equality and the work of Hummer and Szabo 
\cite{jarzy,hs}, Eq. \ref{logP} firmly establishes 
$\ln p^{ness}$ as a NePf of the NESS.  More 
specifically, these previous studies have shown that the 
integration of work (or heat), though itself is not path-independent 
and a function of the state, its LME is a function of state. 
Jarzynski in fact has shown that the LME of work is the free energy 
of the system.  Now Eq.~\ref{logP} shows that the LME of heat 
dissipation is a generalization of Boltzmann's entropy.  
Note that for equilibrium steady-state, $Q_{irr}(t)\equiv 0$ and
Eq.~\ref{logP} becomes a trivial equality. 

	As the celebrated Jarzynski's equality, Eq. \ref{logP} is 
valid for all finite $t$.  We note that Eq.~\ref{logP} does not 
provide a novel way to compute $p^{ness}$ since computing heat 
dissipations, the $q$'s, requires knowing $p^{ness}$ {\it a priori} 
(Fig. 1).  However, if we replace $Q_{irr}(t)$ by the internal 
work $W_{int}(t)=-\sum_{i=1}^n\Delta\mu_{x_{i-1}x_i}^o$
$=-\mu^o_{x_n}+\mu^o_{x_0}$, 
then we have
\begin{equation} 
	k_BT\ln\Big\langle\delta_{X,x_n}
		e^{-W_{int}(t)/k_BT}\Big\rangle^{x_0}
	= \left(\mu_X^o+k_BT\ln p_X(t)\right)
		- \left(\mu_{x_0}^o+k_BT\ln p_{x_0}(0)\right),
\label{logPuseW}
\end{equation}
where $\langle\cdots\rangle^{x_0}$ denotes ensemble average 
over all possible trajectories which start at $x_0$.  
The right-hand-side is the chemical potential difference 
between $X$ at time $t$ and $x_0$ at time zero.  This result 
resemble very much the Jarzynski equality, in which the two end 
states are equilibrium states, while here the two end states are 
two configurations of a NESS.

\vskip 0.5cm\noindent
{\bf NePf based on stochastic macromolecule mechanics of driven
single molecules} 
\vskip 0.3cm 

	The above results for the 3-state driven system can be 
generalized to any single macromolecules in aqueous solution with
stochastic dynamics, either in terms of discrete master equations 
\cite{schn} or continuous Brownian dynamics \cite{leb,hdf3}.  The 
discrete approach is a tradition of chemical kinetics.  For a 
single macromolecule, the kinetics consists of unimolecular reactions 
with first-order rate equations and pseudo-first-order rate 
equations when cofactors are involved.  The continuous Brownian
(Langevin) approach to the dynamics of single macromolecules began with 
Kramer's theory \cite{kramer} and has been extensively studied in
polymer dynamics \cite{doi-edwards}.  We shall develop the general 
theory of NePf based on the Brownian dynamic formalism below. 

	Based on the polymer dynamic approach, we have recently 
developed a nonequilibrium thermodynamic theory for overdamped 
single macromolecules, such as a motor protein, in aqueous solution,
which we called stochastic macromolecular mechanics 
\cite{hdf3,qianjpc02,qqt02}.  We have shown how thermodynamic concepts
such as entropy production rate, heat dissipation rate, and Onsager's 
thermodynamic force are rigorously developed from the dynamic 
equation \cite{hdf3}.  Furthermore, we have introduced mesoscopic, 
trajectory-based fluctuating energy $U_t\equiv U(X_t)$ and 
entropy $\Upsilon_t\equiv -k_B\ln P(X_t)$ where $X_t$ is 
a stochastic trajectory of a macromolecule in its 
$3N$-dimensional phase space and $P(x)$ is the probability 
density.  We observed that in an isothermal equilibrium 
$U_t-T\Upsilon_t \equiv 0$ along a stochastic trajectory
(Recall the above example.)  Beyond equilibrium steady-state, we have 
suggested that $\mu(x)$ $\equiv$ $U(x)-T\Upsilon(x)$ as the 
nonequilibrium chemical potential function for the macromolecule 
in its driven NESS.  Introducing the fluctuating entropy $\Upsilon_t$ 
in terms of the $P(x)$, however, is not completely satisfying 
since it requires the probability density being solved from a 
Fokker-Planck equation {\it a priori}.  Also from an 
experimental standpoint as well as a molecular simulation 
standpoint, this definition is not very useful since it can 
not be obtained along stochastic trajectories $X_t$.
The fluctuating energy in Eq. \ref{logPuseW}, 
however, is computable from a stochastic trajectory
(The statistical accuracy, of course, is an important 
but separated issue.)
	
	In this section we establish the NePf, $-\ln P(x)$, 
and the nonequilibrium chemical potential, $\mu(x)$, through 
FTs-like equalities (Eqs. \ref{logP} and \ref{logPuseW}) in 
terms of stochastic dynamics of a single macromolecule in 
aqueous solution \cite{hdf3} 
\begin{equation}
     \eta dX_t = \vF(X_t)dt + \sqrt{2k_BT\eta}\ dB_t.
\label{sde}
\end{equation}
In Eq. \ref{sde} $\vF(x)$ is decomposed into an internal potential 
part and an external driving part: $\vF(x)$ $=-\nabla U(x)+\vG(x)$. 
$\eta$ is a frictional coefficient and we assume it is a scalar 
for simplicity. $B_t$ is the standard Brownian motion which 
represents the collision between the atoms in the macromolecule 
and the solvent \cite{hdf3,doi-edwards}.  The probability density 
function for $X_t$ is the solution to a Fokker-Planck equation 
\begin{equation}
       \partial P(x,t)/\partial t = -\nabla\cdot\vJ(x,t)
     =(k_BT/\eta)\nabla^2 P -(1/\eta)\nabla\cdot\left(\vF P\right).
\label{fpe}
\end{equation}

	All FTs focus on the work done by various stochastic forces.  
There are three forces in our problem: the total force $\vF$, the 
external force $\vG$, and the Onsager's thermodynamic force $\vPi$ 
$=\vF-k_BT\nabla\ln P(x,t)$ \cite{hdf3}:
\begin{equation}
    W_{tot}(t) = \int_0^t \vF(X_s)\circ dX_s, \hspace{0.3cm}
    W_{ext}(t) = \int_0^t \vG(X_s)\circ dX_s,\hspace{0.3cm}
    Q_{irr}(t) =  \int_0^t \vPi(X_s,s)\circ dX_s.
\label{sf}
\end{equation}
These stochastic integrals are called total work, external work, 
and irreversible heat dissipation/total entropy production, respectively.  
We use `$\circ$' for Stratonovich and `$\cdot$' for Ito integrations.  

	If the macromolecule is in a closed vessel, i.e., $\vG(x)=0$, 
then both $\vF$ and $\vPi$ are conservative forces with potential functions
$U(x)$ and $U(x)-k_BT\ln P(x,t)+Const.$, the internal energy and 
Helmholtz free energy (generalized), respectively \cite{qianre}.  
More importantly, in equilibrium, i.e., the stationary state of the 
closed system,  $P^{eq}(x)\propto$ $e^{-U(x)/k_BT}$, and $\vPi =0$ as 
expected: The Onsager's thermodynamic force is zero.
Then $W_{tot}(t)$ is the fluctuating potential
energy and $Q_{irr}(t)$ is the fluctuating free energy. 
$Q_{irr}(t)\equiv 0$ when the macromolecule is in an equilibrium
\cite{hdf3}. 

	It is easy to show that 
\begin{equation}
           \langle Q_{irr}\rangle(t) =
   \int_0^t ds\int dx\vPi(x)\cdot\vJ(x,s)
	 = \eta\int_0^t ds\int dx P^{-1}\vJ^2(x,s),
\end{equation}
is the mean entropy production which is always non-negative.
\begin{equation}
           \langle W_{tot}\rangle(t) =
   \int_0^tds \int dx \vF(x)\cdot\vJ(x,s),
\end{equation}
however, is the mean heat dissipation which can be negative in 
transient relaxation. $W_{tot}(t)$ being the total heat dissipation 
has been discussed by several authors \cite{sekimoto,hdf2,hdf3}. 
Finally, since
\begin{equation}
         W_{ext}(t) = W_{tot}(t)-W_{int}(t)
		    = W_{tot}(t) + U(X_t)-U(X_0),
\end{equation}
we see that the work done by the external 
force is equal to the heat dissipation plus the increase in 
internal potential energy.  This is the trajectory-based
energy conservation.  In a NESS the internal potential energy 
fluctuations and entropy fluctuations are both stationary, i.e., 
$E\left[U(X_t)\right]$ and $E\left[\ln P(X_t)\right]$ are
constants, hence $\langle W_{ext}\rangle(t)$ $=\langle W_{tot}\rangle(t)$
$=\langle Q_{irr}\rangle(t)$.  The work done by the external force is 
completely transformed into dissipated heat, which precisely 
equals to the isothermal entropy production, in the NESS \cite{bct}.

	In terms of irreversible heat $Q_{irr}(t)$, we now introduce 
\begin{equation}
  R(x,t) = \Big\langle\delta(x-X_t)
		e^{-Q_{irr}(t)/k_BT}\Big\rangle
\label{Rxt}
\end{equation} 
where $\langle\cdots\rangle$ denotes ensemble average over
all possible $X_t$ starts with $R(x,0)=P^{ness}(x)$.  It then can 
be shown mathematically that the following equality holds true for 
all time $t$ (see Appendix A):
\begin{equation}
           \ln P^{ness}(x) = \ln\Big\langle\delta(x-X_t)
		e^{-Q_{irr}(t)/k_BT}\Big\rangle,
\label{logP2}
\end{equation}
which is our main result. It proves Eq. \ref{logP}: 
{\it The trajectory-based log-mean-exponential (LME) of the irreversible
heat dissipation is the NePf ,} as previously suggested by Graham 
\cite{graham4}, Ao \cite{aoping1}, and Qian \cite{qianpre04}.
Note that the ``Boltzmann entropy'' in Eq. \ref{logP2} is defined
through multiple trajectories with initial distribution $P^{ness}$.
Alternatively with arbitrary initial distribution, Eq. 
\ref{logP2} holds only for $t\rightarrow\infty$. 

	We now consider $W_{int}(t)$, the work done by the 
internal potential force $-\nabla U$.  $W_{int}$ is related 
to $W_{tot}$ and and $W_{ext}$ (Eq. \ref{sf}) and its ensemble
average is zero in a NESS. The exponential average
\begin{equation}
  Q(x,t|y) = \Big\langle\delta(x-X_t)
		e^{-\lambda W_{int}(t)/k_BT}\Big\rangle^y
\end{equation} 
where $\langle\cdots\rangle^y$ denotes ensemble average over
all $X_t$ which start at $X_0=y$. Noting that $W_{int}(t)=$
$U(X_0)-U(X_t)$, it is easy to see that 
\begin{equation}
   k_BT\ln Q(x,t|y) = U(x)+k_BT\ln P(x,t|y)-U(y) 
		\equiv \mu(x,t) + \textrm{ const.}
\label{tdcp}
\end{equation}
is, off by a constant, the nonequilibrium chemical potential 
at time $t$.  In other words, {\it the trajectory-based 
LME of the internal potential energy is the 
nonequilibrium chemical potential.}  When $t\rightarrow\infty$, 
we have
\begin{equation}
	k_BT\ln\Big\langle \delta(x-X_t)e^{-W_{int}(t)/k_BT}
	\Big\rangle^y \rightarrow  \mu^{ness}(x)
		\equiv U(x)+k_BT\ln P^{ness}(x)-U(y).
\label{necp}
\end{equation}
Eqs. \ref{tdcp} and \ref{necp} for transients and the NESS of 
open driven systems, our second main result, are parallel to 
Jarzynski's equalities \cite{jarzy,hs} for time-dependent but 
detail-balanced systems. 

	The mathematics behind Eq. \ref{tdcp} is trivial since the 
$W_{int}$ is the work done by the internal potential force.
In fact, to obtain $P(x,t)$, the simplest way is by mathematical
definition $P(x,t|y)$ $=\langle \delta(x-X_t)\rangle^y$.  We emphasize 
that the physically novel result is Eq.~\ref{logP2}, which defines
the $\ln P^{ness}$ as a thermodynamic potential.  The physical meaning 
of Eqs. \ref{tdcp} and \ref{necp}, then, is to establish equalities 
between other thermodynamic potentials and the LME of other types of work 
in an open system.  In this spirit, we have a third relation 
of such (Appendix B):
\begin{equation}
	k_BT\ln\Big\langle \delta(x-X_t)e^{-W_{ext}(t)/k_BT}\Big\rangle^y
	= \mu(y,t) + \textrm{const.}
\label{3rd}
\end{equation}
where the constant is independent of $y$ and $t$; It is a function
of $x$.

\vskip 0.5cm\noindent
{\bf NESS thermodynamics of single driven macromolecules: a theory}
\vskip 0.3cm

	 With the chemical potential $\mu(x,t)$ defined, the 
free energy of a single macromolecule $\Psi(t)$ $\equiv$
$\int \mu(x,t)P(x,t)dx$ has the fundamental property
expected from a nonequilibrium generalization of Helmholtz 
free energy \cite{hdf3}:
\begin{eqnarray}
  \frac{d}{dt}\Psi(t) &=& \int \nabla\mu\cdot\vJ dx 
   = E\left[W_{ext}(t)\right]-e_p 
\nonumber\\
       &=& E\left[W_{ext}(t)\right]
	- h_d - T\frac{d}{dt}S
\label{balance}
\end{eqnarray}
where $e_p$ $=\langle Q_{irr}(t)\rangle/t$
$=\eta\int \|\vJ\|^2P^{-1}dx$ is total entropy production rate,
$h_d$ $=\langle W_{tot}(t)\rangle/t$
$=\int \vF\cdot\vJ dx$ is the heat dissipation rate
\cite{hdf3}, and $S=-k_BT\int P\ln P dx$ is entropy.  The 
first term in (\ref{balance}) is the work done by external work 
-- energy input.  The sum of the first two terms is the energy 
stored in the system -- potential energy increase,
$\Delta U$.  Hence, the relation $\Delta\Psi = \Delta U - T\Delta S$ 
hold true even for transient driven nonequilibrium processes.  
In a NESS, both $d\Psi/dt$ and $dS/dt$ are zero and the energy input are
transformed entirely into dissipated heat. 
Conversely, $\mu^{ness}(x)$ can be written in terms of 
functional differentiation, as in classic thermodynamics: 
\begin{equation}
        \mu^{ness}(x) = \frac{\delta\Psi}{\delta P^{ness}(x)}.
\end{equation}
$\mu(x,t)$ also plays an important role in the transient
dynamic relaxation to a NESS.  One can introduce chemical
potential difference $\mu(x,t)-\mu^{ness}(x)$ and its average 
\begin{equation}
        H(t) = \int P(x,t)\left(\mu(x,t)-\mu^{ness}(x)\right) dx
             = \int P(x,t)\ln
	\left(\frac{P(x,t)}{P^{ness}(x)}\right)dx.
\label{relent}
\end{equation}
This is a Lyapunov function for the stochastic dynamics (Eq. 
\ref{fpe}): $H(t)\ge 0$ and $(d/dt)H(t) \le 0$ \cite{leb,qianre}. 
Therefore, {\it a transient process relaxing to a NESS minimizes,
on average, the chemical potential difference with respect to
its NESS.}

	Hatano and Sasa \cite{sasa} have recently developed a 
theory for steady-state thermodynamics of stochastic systems.
The relation between their work and ours \cite{hdf2,hdf3} is 
as follows.  The key thermodynamic quantities in their theory 
are total heat dissipation $W_{tot}(t)$, housekeeping heat 
dissipation, which is precisely our $Q_{irr}(t)$ in Eq. 
\ref{sf}, and mesoscopic entropy $\Upsilon_t$ = $-k_B\ln P^{ness}(x)$.   
As we have shown, the mean housekeeping heat 
$\langle Q_{irr}\rangle$ is exactly the entropy production rate 
($e_p$) in our work \cite{hdf2,hdf3}, this relation is further 
explored in \cite{seifert2}.  Their irreversible 
force $b(x)$ $=G(x)-\nabla\mu^{ness}(x)$ is exactly $\vPi$, 
the Onsager's thermodynamic force.  We derived an entropy balance 
equation in a NESS $dS/dt$ = $e_p-E[W_{tot}(t)]$ which they
called excess heat $Q_{ex}$.  Their study, however, 
has gone beyond simple NESS systems and their main
result is concerned with time-dependent driven system, i.e,
$\vF$ in (\ref{sde}) is also an explicit function of $t$.
In parallel with \cite{jarzy} and treating $\Upsilon(x,t)$ as
a time-dependent potential energy function, they introduced 
a charging process
\begin{equation}
   Y(t) = \int_0^t \frac{\partial\Upsilon}{\partial s}(X_s, s)ds
     = \frac{1}{T}\int_0^t \frac{\partial 
			(U-\mu^{ness})}{\partial s}(X_s, s)ds
\end{equation} 
and found an interesting generalization of Jarzynski's result to
driven systems.  The mean $\langle Y(t)\rangle$ is a generalization 
of the relative entropy $H(t)$ in (\ref{relent}).  The present work, 
however, is limited to time-independent $\vF(x)$ and is concerned 
with establishing the concept of NePf and nonequilibrium chemical 
potential in terms of the LME of certain type
of work in a nonequilibrium open system.

\vskip 0.5cm\noindent
{\bf Discussions}
\vskip 0.3cm

	{\it Stochastic NESS circuit theory.}  
An analogy between the nonequilibrium steady-state 
thermodynamics and electrical network theory exists.  
$\mu$ represents potentials, $\nabla\mu$ represents potential 
gradient, $\vJ$ represents flux, and $\nabla\cdot\vJ=0$ is 
Kirchhoff's current law.  Tellegen's theorem then says that 
$\nabla\mu$, is perpendicular to $\vJ$ in a NESS:
\begin{equation}
        \int \nabla\mu(x)\cdot\vJ(x)dx = 
        -\int \mu(x)\nabla\cdot\vJ(x)dx = 0.
\end{equation}
In the case of a system being driven at its boundary,
$\vF(x)$ $=-\nabla U(x)$ for $x\in\Omega$ and the 
nonequilibrium driving force is only
imposed on the boundary $\partial\Omega$:
\begin{equation}
        \int_{\Omega} \nabla\mu(x)\cdot\vJ(x)dx = 
       \int_{\partial\Omega} \mu(x)\vJ(x)\cdot d\vS
\end{equation}
is analogous to Kirchhoff's potential law.  
For $x\in\partial\Omega$, $\vJ\neq 0$ is called flux injection
and $\mu(x)\neq 0$ is called potential clamping 
\cite{bct,beard1,beard2}.  If there is neither flux injection 
nor potential clamping, then the system ($x\in\Omega$) is 
closed and it reaches equilibrium with $\vJ=\nabla\mu=0$. 

\vskip 0.3cm

	{\it Graham-Haken decomposition}.  In the present
work, the decomposition of nonconservative force $\vF(x)$ 
= $-\nabla U(x)+\vG(x)$ is based on the physical reality
of open and closed systems: Any open system has a
unique corresponding closed system with internal potential
$U(x)$.  Knowing this correspondence greatly enhances 
the thermodynamic understanding of the open system 
(see \cite{qianpre04,crooks}).   
Mathematically, however, Helmholtz-Hodge 
decomposition $\vF=-\nabla\phi+\gamma$ is not unique.  
Graham and Haken \cite{graham1} have shown that if one 
chooses $\phi=-k_BT\ln P^{ness}(x)$, then $\gamma$ 
satisfies $\nabla\cdot\gamma-\nabla\phi\cdot\gamma$ $=0$.
In terms of this particular decomposition, our $\mu(x)$ in 
Eq. \ref{tdcp} $\equiv$ const., and more importantly, 
$Q^{gh}(x,t)$ = $\int Q(x,t|y)dy$ $\equiv 1$.  That is, the 
exponential average of Graham-Haken ``internal work'', over
an initially uniform distribution, is 1 for any finite time $t$. 
The significance of this result is that one, having experimental
or simulated realizations of many trajectories of $X_t$ with
finite time duration $X_t$ $(0\le t\le t^*$), can devise an 
inverse algorithm to construct $\phi(x)$ such that
\begin{equation}
  k_BT\ln\Big\langle\delta(x-X_t)e^{(\phi(X_t)-\phi(X_0))/k_BT} 
  \Big\rangle \equiv  0.          
\end{equation}
Finally, we notice that if one chooses the Graham-Haken decomposition,	
then by the result in footnote 3, we have
\begin{equation}
\langle\delta(y-X_t)e^{-\int_0^t \gamma(X_s)\circ dX_s}\rangle^x
=
\langle\delta(x-X_t)e^{\int_0^t \nabla\phi(X_s)\circ dX_s}\rangle^y 
\rightarrow P^{ness}(y)
\end{equation}
in the limit of $t\rightarrow\infty$.

\vskip 0.3cm

	{\it Umbrella sampling in stochastic dynamics}.  
Umbrella sampling is an important method in molecular
simulations \cite{chandler}.  Our result on nonequilibrium
chemical potential also has an intimate relation to 
a more general mathematical result which has 
applications for umbrella sampling.  Consider an
arbitrary function $V(x)$ and integral
$I_V(t)$ = $-\int_0^t \nabla V(X_s)\circ dX_s$, then 
\begin{equation}
    \Big\langle\Big\langle\delta(x-X_s)e^{-I_V(t)/k_BT}\Big\rangle^y
	e^{(V(y)-V(x))/k_BT}\Big\rangle
       = P^{ness}(x)
\end{equation}
in which the inner $\langle\cdots\rangle^y$ is an averaging over all 
$X_t$ with $X_0=y$ and the outer $\langle\cdots\rangle$ is an
averaging over all $X_0$ with the stationary distribution of $X$.  
The result is the stationary distribution which can be obtained 
from sampling of $X_t$ weighted toward those trajectories which move 
along with the gradient of $V(x)$. 

\vskip 0.3cm

	{\it Conclusions.}  Open, driven chemical and molecular 
systems far from chemical equilibrium have been studied from 
several different points of view in statistical physics.  In terms of 
stochastic processes, these investigations include 
``nonequilibrium potentials'' \cite{qianpre04,graham4,aoping1}, 
``dissipative structures'' \cite{Prig},  ``cycle kinetics'' 
\cite{qianjpcm05,hill1,hill2}, ``Brownian motors'' 
\cite{JAP,qianjpcm05}, ``stochastic resonance'' \cite{stores}, and 
``fluctuation theorems''\cite{ft1,ft2,crooks,seifert}. 
The result we obtain in this work, in terms of a Jarzynski-type
equality, clearly demonstrates an intimate connection between
the fluctuation theorems and the nonequilibrium potentials.
Combining with previous work which has shown relations
between Brownian motors and cycle kinetics \cite{qianprl98},
between cycle kinetics and fluctuation theorems \cite{hdf2},
between Brownian motors and excess entropy \cite{hdf4},
and between stochastic resonance and cycle kinetics \cite{qianprl00},
a consistent and comprehensive statistical thermodynamic theory 
for single macromolecules far from equilibrium emerges. 
Finally but not least, a variational principle for NESS in terms 
of LME of heat dissipation seems to be in sight.

I thank P. Ao, Z.Q. Chen, R. Fox, G. Gallavotti, R. Graham, C. 
Jarzynski, S.-i. Sasa, J.M. Schurr, U. Seifert, Z.F. Shao, and 
X.S. Xie for helpful discussions.

\vskip 0.5cm\noindent
{\bf Appendices} 
\vskip 0.3cm 

	In stochastic mathematics, Langevin equations such 
as in Eq.~\ref{sde} are known as stochastic differential 
equations \cite{oks}.  Because the random force 
$dB_t$, a solution to the equation is stochastic and
has to be characterized in terms of a probability density
function $P(x,t)$ which satisfies the corresponding
Eq.~\ref{fpe}.  For arbitrary force $\vK(x,t)$, let 
$\phi(x,t)=$ $\big\langle\delta(x-X_t)
e^{-\int_0^t\vK(X_t,t)\circ dX_t/k_BT}\big\rangle$ 
where $\langle\cdots\rangle$ denotes ensemble average
and let $\widehat{\phi}(t,x)=$ 
$\big\langle e^{-\int_0^t\vK(X_t,t)\circ dX_t/k_BT}\big\rangle^x$ 
where $\langle\cdots\rangle^x$ denotes ensemble average
over all possible trajectories (solutions to Eq.~\ref{sde}) 
which start at $x$,  the Cameron-Martin-Girsanov formula 
\cite{oks} states that $\widehat{\phi}(t,x)$ satisfies the 
equation $\partial\widehat{\phi}/\partial t$ = 
${\mathcal L}^*\widehat{\phi}$ and
$\phi(x,t)$ satisfies the equation $\partial{\phi}/\partial t$ 
= ${\mathcal L}\phi$, where the elliptic operator ${\mathcal L}=$
$(k_BT/\eta)\nabla^2-(1/\eta)[\vF\cdot\nabla+\nabla\cdot\vF
-2\vK\cdot\nabla-\nabla\cdot\vK-(1/k_BT)\vK\cdot(\vK-\vF)]$,
and ${\mathcal L}^*$ is the adjoint of ${\mathcal L}$.

	Now for arbitrary scale $q(x,t)$, let 
$\psi(x,t)=$ $\big\langle\delta(x-X_t)
e^{-\int_0^t q(X_s,s)ds/k_BT}\big\rangle$ and
$\widehat{\psi}(t,x)=$ 
$\big\langle e^{-\int_0^t q(X_s,s)ds/k_BT}\big\rangle^x$,
the Feymann-Kac formula \cite{oks}  states that 
$\widehat{\psi}(t,x)$ satisfies the 
equation $\partial\widehat{\psi}/\partial t$ = 
${\mathcal L}^*\widehat{\psi}$ and
$\psi(x,t)$ satisfies the equation $\partial{\psi}/\partial t$ 
= ${\mathcal L}\psi$, where the operator ${\mathcal L}=$
$(k_BT/\eta)\nabla^2-(1/\eta)[\vF\cdot\nabla+\nabla\cdot\vF]
-(1/k_BT)q(x,t)$.  These two adjoint relations are
behind the work of Jarzynski \cite{jarzy}
and that of Hummer and Szabo \cite{hs}.  In those studies,
the work $W(t)$ is defined through an integral
$\int_0^t F(X_s,s)v(s)ds$ where $F(x,s)$ is a time-dependent
external force and $v(s)$ is a deterministic rate.  Hence
Feymann-Kac, instead of Girsanov, formula was used. 

\vskip 0.3cm

	{\it Appendix A.} $R(x,t)=$ 
$\big\langle\delta(x-X_t)e^{-Q_{irr}(t)/k_BT}\big\rangle$ with
$R(x,0)=P^{ness}(x)$.  That is the above $\vK = \vPi$ in 
Eq.~\ref{sf}.  Hence $R(x,t)$ satisfies the partial differential 
equation $\partial R(x,t)/\partial t$ $={\mathcal L}R(x,t)$ where 
\begin{equation}
   {\mathcal L} = \frac{k_BT}{\eta}\nabla^2 -
	\frac{1}{\eta}\left[\vF\cdot\nabla+(\nabla\cdot\vF)
         -2\vPi\cdot \left(\nabla-\nabla\ln P^{ness}\right)\right].
\end{equation}
It is easy to verify that $P^{ness}(x)$ is the solution to this 
equation for all time $t$.   Hence we arrive at Eq.~\ref{logP2}.

\vskip 0.3cm

	{\it Appendix B.}  If the $\vF$ in 
Eq.~\ref{sde} has $\vF=\vF_1+\vF_2$, then applying the
above result  we have
$\langle\delta(x-X_t)e^{-\int_0^t \vF_1(X_s)\circ dX_s}\rangle^y$ $=$
$\langle\delta(y-X_t)e^{-\int_0^t \vF_2(X_s)\circ dX_s}\rangle^x$.
A direct consequence of this result is a symmetry relation
$\langle\delta(x-X_t)e^{-\lambda W_{tot}(t)}\rangle^y$ $=$
$\langle\delta(y-X_t)e^{-(1-\lambda)W_{tot}(t)}\rangle^x$.
This is an alternative statement of the FTs.

	Now let $Z(y,t)=\langle\delta(x-X_t)e^{-W_{ext}(t)/k_BT}\rangle^y$.  
As a function of $y$, it satisfies the Kolmogorov backward 
equation $\partial Z(y,t)/\partial t$ $={\mathcal M}Z(y,t)$ where 
the operator ${\mathcal M}=$ 
$(k_BT/\eta)\nabla^2-(1/\eta)[\vF\cdot\nabla
+(\nabla\cdot\vF)+2(\nabla U)\cdot\nabla+\nabla^2 U]
+(1/\eta k_BT)\nabla U\cdot\vG$ 
$=e^{U/k_BT}{\mathcal L}_0 e^{-U/k_BT}$. 
${\mathcal L}_0$ is the right-hand-side of Eq.~\ref{fpe}. 
Because $W_{ext}+W_{int}=W_{tot}$, we arrive at Eq. \ref{3rd}.

\end{document}